\documentclass[conference]{IEEEtran}
\IEEEoverridecommandlockouts
\usepackage{optidef}
\usepackage{authblk}
\usepackage{cite}
\usepackage{booktabs}
\usepackage{tabu}
\usepackage{amsmath,amssymb,amsfonts}
\usepackage[hang,flushmargin]{footmisc}
\usepackage{multirow}
\usepackage{float}
\usepackage{textcomp}
\usepackage{lscape}
\usepackage{url}
\usepackage{enumerate}
\usepackage{caption}
\usepackage{subcaption}
\usepackage{algpseudocode}
\usepackage{graphicx}
\usepackage{textcomp}
\usepackage{xcolor}
\usepackage{array}
\usepackage{mathtools,algorithm}

\usepackage{enumitem}

\newcommand{\CASE}[1]{\STATE \textbf{case} #1\textbf{:} \begin{ALC@g}}
\newcommand{\ENDCASE}{\end{ALC@g}}

\newcommand{\DEFAULT}{\STATE \textbf{default:} \begin{ALC@g}}
\newcommand{\ENDDEFAULT}{\end{ALC@g}}
\newcommand{\DEFAULTLINE}[1]{\STATE \textbf{default:} }
\newcolumntype{L}[1]{>{\raggedright\let\newline\\\arraybackslash\hspace{0pt}}m{#1}}
\newcolumntype{C}[1]{>{\centering\let\newline\\\arraybackslash\hspace{0pt}}m{#1}}
\newcolumntype{R}[1]{>{\raggedleft\let\newline\\\arraybackslash\hspace{0pt}}m{#1}}
\def\BibTeX{{\rm B\kern-.05em{\sc i\kern-.025em b}\kern-.08em
    T\kern-.1667em\lower.7ex\hbox{E}\kern-.125emX}}

\begin{document}

\title{Dynamic Routing and Spectrum Assignment based on the Availability of Consecutive Sub-channels in Flexible-grid Optical Networks}

\author[ ]{Varsha Lohani}
\author[ ]{Anjali Sharma}
\author[ ]{Yatindra Nath Singh} 

\affil[  ]{Department of Electrical Engineering, Indian Institute of Technology Kanpur, Kanpur, India}

\affil[  ]{\textit {lohani.varsha7@gmail.com*, anjalienix05@gmail.com and ynsingh@iitk.ac.in}}

\maketitle

\begin{abstract}
Variable bandwidth channels can be created in Flexible Grid Optical Networks using Optical Orthogonal Frequency Division Multiplexing (O-OFDM). This allows more efficient use of spectrum by allocating integral multiple of basic bandwidth slot to the lightpath requests. In these networks, the constraint of keeping all the allocated slots together is added when deciding the routes for the requests. This constraint is called the contiguity constraint, which makes the routing and spectrum arrangement algorithms more challenging. In any network, the lightpath requests will arrive and depart dynamically and invariably lead to spectrum fragmentation. Hence network will have to reduce the maximum possible utilization as well as increased blocking probability. In this paper, we have presented an improvised Routing and Spectrum Assignment (RSA) algorithm using consecutive spectrum slots that leads to lesser fragmentation. It is evident from the results that the presented RSA algorithm uses adaptive parameters to reduce the blocking probability and fragmentation compared to the other algorithms reported in the recent past.

\end{abstract}

\section{Introduction}

With the innovations continuously improving the performance of network endpoint devices, there is a need to continuously improve communication network capacities to meet the resulting demands. All-Optical networks are used to provide the needed increased capacity. The signal that traverses from a source to destination node in these networks, remains in the optical domain. These networks contain routing nodes interconnected by optical fiber links. The resources used in these links can be either fixed-width wavelength slots or flexible spectrum slots. Fixed width wavelength slots are generally based on the Dense Wavelength Division Multiplexing technique. The bandwidth of each channel within the network link, is fixed as either 50 GHz or 100 GHz according to ITU-T G.694.1 \cite{itut} specifications. High data-rate connection cannot established due to limit posed by the slot width. At the same time, a very low data-rate lightpath requests will lead to bandwidth wastage. Jinno \textit{et al.} \cite{EON1} proposed the spectrum sliced elastic optical path networks to reduce the above problems. 

The OFDM-enabled technology facilitates the allocation of a group of frequency slots (FSs) to each lightpath request according to the required bit rate. They provide a scalable network architecture using Bandwidth Variable-Wavelength Cross-connects (BV-WXC) and Bandwidth Variable Transponders (BVT). An overview of Flexible-grid Optical Network and the related design issues can be found in \cite{RSA4} and \cite{RSA9}. 

Use of Optical- Orthogonal Frequency Division Multiplexing (O-OFDM), allows the further reduction of the channel size to 12.5 GHz further improving the granularity of bandwidth allocation. The O-OFDM adds flexibility to the optical network by permitting the variable connection bandwidth in multiples of 12.5 GHz. With flexibility, multiple adjacent channels can be used together to accommodate high data-rate connection demands. Such networks are called Flexible-grid Optical Networks\footnote{Optical Orthogonal Frequency Division Multiplexing based Elastic Optical Networks}.

One of the research problems in optical networks is Routing and Resource\footnote{fixed wavelength slots or flexible spectrum slots} Assignment (RRA). The routing of a lightpath request implies finding an end-to-end path between source and destination nodes using a suitable algorithm. Resource assignment implies finding out the link resources  for the path setup request while following the relevant constraints. Routing and Resource Assignment for a lightpath request is generally an NP-hard problem. In real scenario, the RRA problem is sub-divided depending upon the type of multiplexing and switching techniques. It is called Routing and Wavelength Assignment (RWA) \cite{rama} in Fixed-grid networks\footnote{Wavelength Division Multiplexing based Optical Networks} and Routing and Spectrum Assignment (RSA) in Flexible-grid networks. 

This paper aims to improve RSA algorithms for flexible grid optical networks to lower blocking probability such that more connection requests can be catered to with reduced computational complexity. However, due to spectrum elasticity, the Flexible-grid also adds contiguity constraints to the RSA problem. This additional constraint and the fluctuating traffic results in the fragmentation of the spectrum. As a consequence, there is an increase in the blocking of lightpath requests. 

We intend to minimize the spectrum fragmentation in the network links and hence the blocking probability of lightpath requests.

The rest of the paper is organised as follows. The basic concept of Routing and Spectrum Allocation is explained through examples in Section II. In Section III, the problem being investigated in this paper is stated. The proposed solution is mentioned in Section IV. In Section V, network definition, tackling unequal capacity on edges, and different proposed RSA algorithms have been presented. Numerical results are shown for two example networks in Section IV, where different network settings are used to understand the performance of the proposed algorithms. 

\section{Routing and Spectrum Assignment (RSA)}

Let's consider an \textit{N} node optical network and suppose each node has \textit{N - 1} transceivers\footnote{transmitters (lasers) and receivers (photodetectors)}. Each node pair can be connected by a dedicated lightpath if there are adequate resources. Nevertheless, such a solution will result in high implementation costs and may not utilize the available spectrum slots efficiently. The traffic demand should be routed using minimum transceivers and wavelengths as it will need switches with a lesser number of ports\cite{RWA11} thus reducing the network cost. The objective of RSA is to find routes over these nodes and allocate spectrum resources to the lightpath requests while using minimum resources and accommodating maximum lightpath requests. 

\begin{figure}
                \centering
                \includegraphics[width=\linewidth]{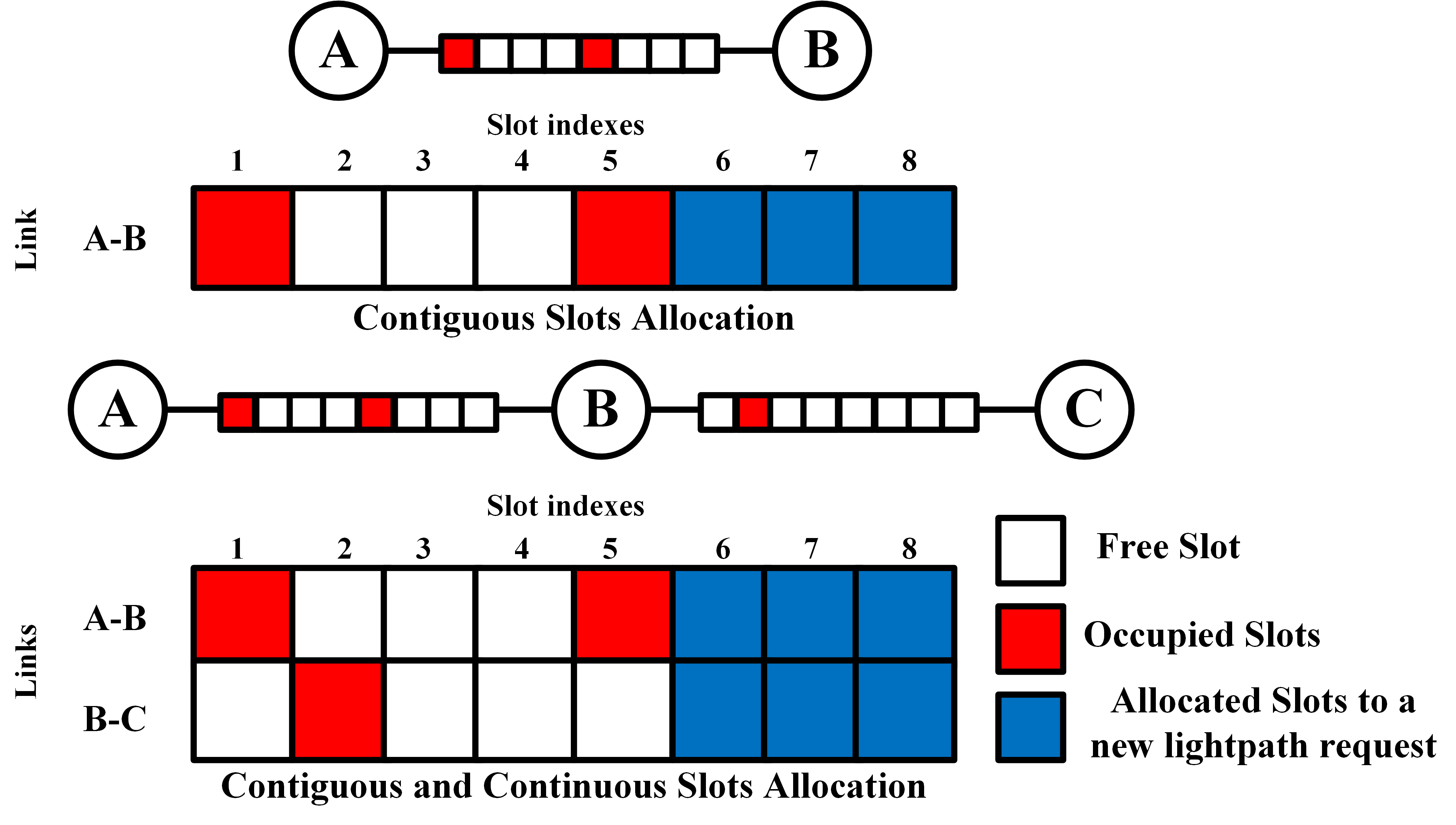}
                \caption{An example of constraints used for Routing and Spectrum Allocation of lightpath request.}
                \label{fig:cc}
            \end{figure}

In RSA, there are constraints for spectrum assignment, which need to be satisfied as explained with the help of Figure \ref{fig:cc}. A lightpath request for three slots connection assignment from source node \textit{A} to destination node \textit{C}.  \textbf{Spectrum Contiguity Constraint:} The assigned spectrum slots indexes 2, 3, 4 or 6, 7, 8  for link A-B and 3, 4, 5 or 4, 5, 6 or 5, 6, 7 or 6, 7, 8 can be used for link B-C as the three slots in possible group share boundary with each other. This is contiguity constraint. \textbf{Spectrum Continuity Constraint:} The assigned continuous spectrum slots indexes 6, 7, 8 for a lightpath request A-B-C is same throughout the path. \textbf{Non-Overlapping  Constraint:} The assigned indexes to different requests cannot overlap with one other. It is consequence of capability of a slot to carry one signal at a time. Therefore, the slot indexes 6, 7, and 8 are assigned as it satisfies the requirements.

Finding a suitable RSA algorithm lies at the core of the Flexible-grid optical networks design. It is supposed to achieve efficient spectrum utilization while accommodating maximum number of lightpath requests.  There exist various RSA algorithms in the literature \cite{RSA3, RSA5, RSA6, RSA7, RSA8, RSA10}, which envisage to reduce the blocking of arriving lightpath requests. In these papers, various heuristics for RSA have been proposed to reduce the blocking of lightpath requests. The problem has been attempted using many methods like employing different routing schemes, devising spectrum assignment algorithms based on the cost of paths, and distance-adaptive modulation level manipulation. In \cite{RSA2}, the modulation format as another dimension for manipulating the bandwidth in different sections of a lightpath, is introduced in the Elastic optical network. The authors in \cite{c8} proposed heuristics for Dynamic routing and spectrum (re)allocation, where the allocated lightpath requests are re-allocated the optical spectrum to make room for new lightpath requests. Few research works \cite{RSA7}, \cite{param} also considered multipath routing for RSA in EON to accommodate more lightpath requests or to reduce the blocking probability. A fixed parameter, e.g., distance (in km), hops, etc., is used for deciding the optimal route in most of these works.  In the current work, we are considering adaptive parameters (which change with network conditions) for RSA. The spectrum resources are allotted at the connection setup time and released only when the connection is dismantled. 

\begin{figure}
    \centering
    \includegraphics[width=\linewidth]{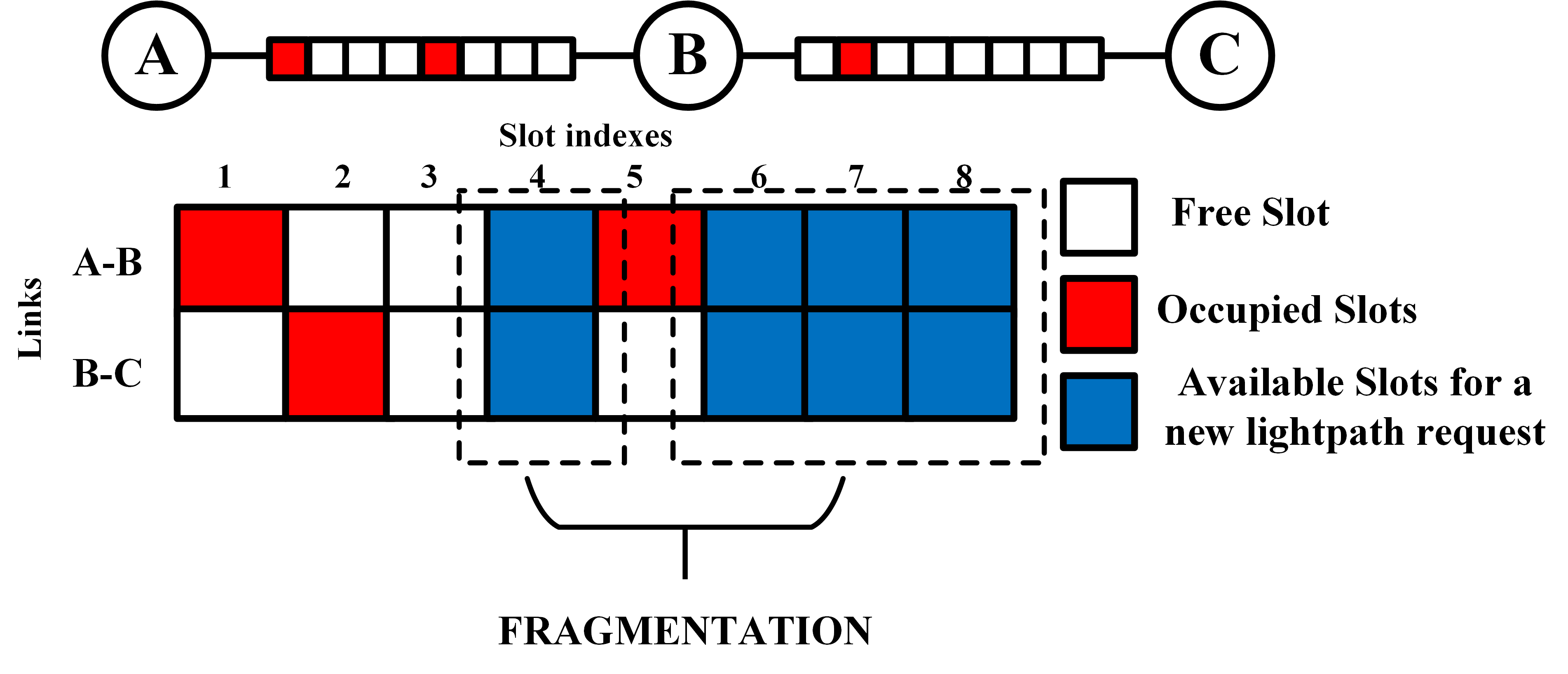}
    \caption{Spectrum Fragmentation example.}
    \label{fig:frag}
\end{figure}

One of the problems due to the spectrum assignment constraints is \textbf{Spectrum Fragmentation}. It can be explained with the help of an example shown in Figure \ref{fig:frag}. A lightpath request for four slots arrives from the source node \textit{A} to the destination node \textit{C}. Although the four slots are available but they cannot be allocated, as we have to follow spectrum assignment constraints (Figure \ref{fig:cc}). It leads to the blocking of the lightpath request due to fragmentation. If spectrum converters are present at each node, the fragmentation can be mitigated to some extent, but converters are expensive to deploy. 

Chatterjee et al. \cite{RSA11}, have extensively studied the various forms of fragmentation. The survey paper presents various types of fragmentation metrics and de-fragmentation strategies. They also discussed various RSA strategies based on fragmentation metrics without any de-fragmentation procedures. Also, numerous de-fragmentation algorithms have been proposed to reduce the spectrum's fragmentation periodically. In \cite{RSA12}, the authors have presented a path-based method to calculate fragmentation level and then to employ it in RSA decision making. \cite{RSA13, RSA14} further explored fragmentation-aware routing and spectrum allocation, considering both contiguity and continuity aspects. Basically, various parameters of the spectrum in the network links are observed, and based on their state; the routing decision is made. These parameters may or may not directly affect the fragmentation level. The contiguity of the spectrum slices is one of the most prominent indicators of fragmentation level. Therefore, it is also possible to use suitable RSA algorithms to minimize contiguity fragmentation during the network operation. The de-fragmentation procedure, whenever invoked, disrupts the existing traffic in the network for the period required for reconfiguration, which leads to another inefficiency in network performance.

% The RSA problem can be further classified into \textit{offline RSA} where the lightpath requests are known in advance and \textit{online RSA} in which the lightpath requests arrive and released as time progresses. The later scenario is more realistic in nature. In this paper, we are considering online RSA without any periodic de-fragmentation. The RSA problem is further sub-classified based on whether the static or adaptive parameters are considered for routing. The static parameters for routing are independent of the changes within the network; while the adaptive parameters for routing change continuously with the change in the network conditions. 

% In this paper, our primary focus is on Routing and Spectrum Assignment for flexible grid optical networks based on adaptive parameter. We propose RSA algorithms to minimize the spectrum fragmentation in the network links and lightpath requests blocking.	

\section{Problem Statement}

The RSA algorithms can be further classified into \textit{offline RSA} where the lightpath requests are known in advance and \textit{online RSA} in which the lightpath requests arrive and released as time progresses. The later scenario is more realistic in nature. In this paper, we are considering online RSA without any periodic de-fragmentation. The RSA problem is further sub-classified based on whether the static or adaptive parameters are considered for routing. The static parameters for routing are independent of the changes within the network; while the adaptive parameters for routing change continuously with the changes in the network conditions. Many of the research works discussed in Section II consider static parameters such as fixed distance and hops\footnote{independent of the network conditions} for routing of lightpath requests from source to destination. Some of them consider the adaptive parameters also for Routing and Spectrum Assignment. \cite{cost1, cost2, cost3} use relative cost parameter (e.g. link congestion) which changes with the network status while performing Routing and Spectrum Assignment for incoming lightpath requests. For various cases, the algorithms have been analyzed. In essence, the algorithms select the route/spectrum set with the least relative cost. Y. Zhou \textit{et. al} \cite{linkstate} computes the link-state based on Chromatic Dispersion and Optical Signal to Noise Ratio (OSNR). This link-state is used for routing purposes. Another adaptive parameter that has been discussed in the literature is crosstalk \cite{ct1, ct2, ct3}. The crosstalk occurs within the multi-core fibres and has a bearing on Routing, Core and Spectrum Assignment in Space Division Multiplexed Elastic Optical Networks. In \cite{ct4}, authors cover the problem of spectrum de-fragmentation in crosstalk aware RCSA. 

\section{Proposed Solution}
One of the adaptive parameters that can be effective for RSA is consecutive spectrum slots present on each network link. One can use this parameter for routing purposes in conjunction with the works mentioned earlier (Section III). In the present work, our objective is to find suitable route and spectrum slots for the incoming lightpath requests using an adaptive parameter called link spectrum consecutiveness. At the same time, we are minimizing the fragmentation within the network without performing any de-fragmentation strategies, so that we can accommodate the maximum number of requests. We are not using any de-fragmentation strategy as it causes interruption of active lightpath requests. 

In this paper, we are considering a dynamic traffic scenario and adaptive routing and spectrum assignment. The same problem can be extended for Routing, Core and Spectrum Assignment. However, we are not considering multiple cores within the fibers in the current work. We are attempting to utilize spectrum as efficiently as possible for single-core Flexible grid Optical Networks. Therefore, we assume no crosstalk.

We propose new RSA algorithms to minimize the spectrum fragmentation in the network links and hence the blocking probability of arriving lightpath requests.

\section{Routing and Spectrum Assignment based on the Availability of Consecutive Slots}
\subsection{Network Model and Notations Used}
\begin{itemize}
    \item $G(V,E, \{ \Delta_{e} \})$: We represent an optical network as a graph \textit{G(V, E)} where \textit{G} is defined as a set of optical vertices\footnote{nodes} \textit{V}, indexed by \textit{v} and set of optical fiber edges\footnote{links} \textit{E}, indexed by \textit{e}. Each edge is connected to a pair of vertices e.g., \textit{$(i, j) \in E$}, where $i$ and $j \in V$. Each edge $\textit{e}\in \textit{E}$ has usable bandwidth, $B_{e}$. The $B_{e}$ is partitioned into multiple spectrum slots in order to be used efficiently. We define a bitmap, $\Delta_{e}$ (sequence of 1s and 0s) on an edge $e \in E$ to model the availability status of the spectrum slots. The cardinality of the possible spectrum slots on an edge \textit{e}, $|\Delta_{e}|$  is represented by
    \begin{equation}
    |\Delta_{e}| = \Bigl \lfloor{\dfrac{\text{Total Usable Bandwidth}(B_{e})}{\text{Grid Size}}}\Bigr \rfloor.
    \label{eqn0}
    \end{equation}
    Suppose each edge has same set (\textit{F}) of spectrum slots ($f_s$), then the bitmap for an edge can be represented as $\Delta_{e} = [f_1 ,....,f_{|F|}]$ where $f_s$ can be either 0 or 1 depending on the condition whether the $s^{th}$ frequency slot is busy or free respectively. The bit $s$ in $\Delta_{e}$ is represented as

        \begin{equation}
            \Delta_{e}[s] = \begin{cases}
            1, &\text{if $s^{th}$ slot on edge \textit{e} is free,}\\
            0, &\text{if $s^{th}$ slot on edge \textit{e} is occupied.}
        \end{cases}
        \label{eqn1}
        \end{equation}

The slot availability status can be easily propagated for a destination through the neighbours by simply performing AND operation of the current status with the slot availability status in links to neighbour.

    \item $LR(s,d,| \Delta^{r} |, k)$ is a Lightpath Request where $s$ is the source node, $s \in V$, and $d$ is the destination node, $d \in V$. In a network \textit{G(V, E)}, if a lightpath request arrives with a required bandwidth of $B^{r}$ for an s-d pair, then the cardinality of the required contiguous spectrum slots on any choosen path is computed with the eq. \ref{eqn5}.
    \begin{equation}
        |\Delta^{r}| = \Bigl \lceil{\dfrac{\textrm{Required Bandwidth}(B^{r})}{\text{Grid Size} * m}}\Bigr \rceil + \Bigl \lceil{\dfrac{\text{GB}}{\text{Grid Size}}}\Bigr \rceil.
    \label{eqn5}
    \end{equation}
    Here m is the number of bits per symbol (modulation level) used depending on the path length of the lightpath. In this paper, we are using Grid size of 12.5 GHz and BPSK modulation format i.e., the modulation level $m = 1$. We are also considering an additional guard band (GB) such that no two lightpaths interfere if they are placed next to each other. The maximum number of paths to be computed by RSA is represented by k, a positive integer. 

    \item $\Delta_{p}$ is the bitmap of available spectrum slots in path $p$ from $s$ to $d$. The availability bitmap for a path $p$, $\Delta_{p}$ can be achieved by intersecting or Bit-wise ANDing of bit maps on all the constituent edges of path $p$ 

\begin{equation}
 \Delta_{p} = \{\Delta_p[l]\} = \{ \bigcap\limits^L_{i=1} \Delta_{e_i}[l] \}.
 \label{eqn3}
\end{equation}
    \item $(u, v) \in E$ is the edge joining the pair of vertices (nodes), where $u \in V$ is the starting (head) node and $v \in V$ is the ending (tail) node.
\end{itemize}

\subsection{Cases of Equal and Unequal Capacities on the network links}

Suppose the fibers are deployed between the pair of nodes in a network at different time instants, the bandwidth capacity of some of the fiber links/edges may be different, subject to technological advancements at the time of deployment. Let the bitmap of edge 1 be $\Delta_{e_1}$ and edge 2 be $\Delta_{e_2}$ and suppose $|\Delta_{e_1}| >|\Delta_{e_2}|$. This is the case of non-uniform bandwidth, hence, the bitmap sequence, on the edges of the network. The $(|\Delta_{e_1}|-|\Delta_{e_2}|)$ slots are zero-padded in $\Delta_{e_2}$ to make uniform size of bitmap sequences for all the links in the network. However, the zero-padded bit sequence remains unavailable for use by lightpath requests independently of the network conditions. Now, $\Delta_{e}[s] = 0$ in eq. \ref{eqn1}, when $s^{th}$ slot on edge \textit{e} is occupied or unavailable.

Eq. \ref{eqn3} holds for both the cases of uniform and non-uniform bandwidth provisioning on the edges in the network. Here, $L$ is the number of hops for path $p$.

\begin{figure}
\begin{subfigure}{0.5\textwidth}
    \centering
    \includegraphics[width=\linewidth]{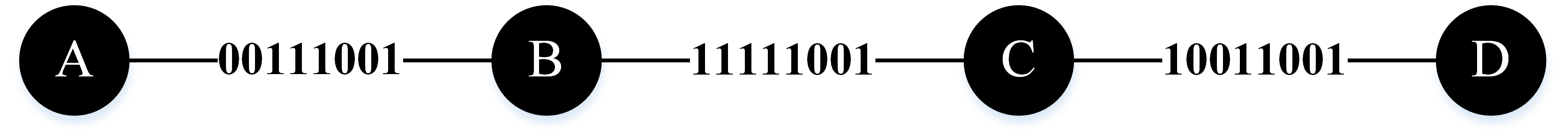}
    \caption{Uniform spectrum slots on links, and }
    \label{fig:eg1}
\end{subfigure}
\begin{subfigure}{0.5\textwidth}
    \centering
    \includegraphics[width=\linewidth]{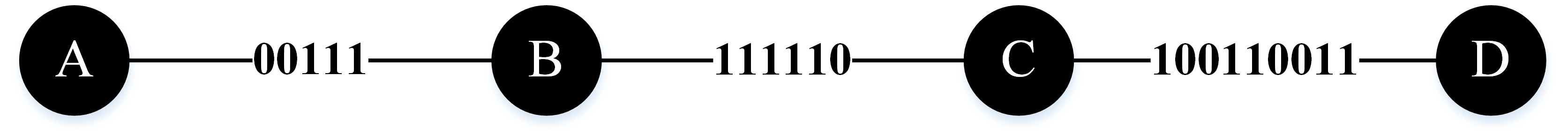}
    \caption{Non-uniform spectrum slots on links.}
    \label{fig:eg2}
\end{subfigure}
\caption{Examples show a lightpath request with three links and their spectrum status.}
\label{fig:e}
\end{figure}

\begin{enumerate}
	\item Consider a uniform network with three link lightpath  ($p$) \textit{A-B-C-D} as shown in Figure \ref{fig:eg1}. Each link on a path has 8 slots. The slot values are represented with 1s and 0s based on the availability status. The value of $\Delta_{e_1}$ for edge 1 is 00111001, $\Delta_{e_2}$ for edge 2 is 11111001, and $\Delta_{e_3}$ for edge 3 is 10011001. Therefore, 
\begin{equation*}
 \Delta_{p} = \{ \bigcap\limits^3_{i=1} \Delta_{e_i} \} = 00011001.
 %\label{eqn4}
\end{equation*}
	\item Now consider the case when the bandwidth on the links is different as shown in Figure \ref{fig:eg2}. The values of $\Delta_{e_1}$ for edge 1 is 00111, $\Delta_{e_2}$ for edge 2 is 111110, and $\Delta_{e_3}$ for edge 3 is 100110011. The cardinality of available slots on edge 1, $|\Delta_{e_1}|$, is 5 on edge 2, $|\Delta_{e_2}|$, is 6, and on edge 3, $|\Delta_{e_3}|$, is 9 i.e., $|\Delta_{e_1}| < |\Delta_{e_2}| < |\Delta_{e_3}|$. The cardinality of edge 3 is the highest so there is no zero-padding for it; however, zero-padding is needed in edge 1 and edge 2. After zero-padding, the cardinality $|\Delta_{e_2}|$ changes to $|\Delta_{e_2}| + (|\Delta_{e_3}| - |\Delta_{e_2}|)$, and $|\Delta_{e_1}|$ changes to $|\Delta_{e_1}| + (|\Delta_{e_3}| - |\Delta_{e_1}|)$. The difference part is zero-padded to the edge 1 and edge 2 slots. Therefore, now the  $\Delta_{e_1}$ for edge 1 is 001110000, and $\Delta_{e_2}$ for edge 2 is 111110000, and hence

\begin{equation*}
 \Delta_{p} = \{ \bigcap\limits^3_{i=1} \Delta_{e_i} \} = 000110000.
 %\label{eqn4}
\end{equation*}
\end{enumerate}

\subsection{Proposed Algorithms}
In this paper, we proposed three algorithms for the joint optimization of Routing and Spectrum Assignment, which use the availability of consecutive spectrum slots for routing. If a lightpath with required slots is available, the spectrum slots are allocated to the request and lightpath is established; otherwise, the request will be blocked.

\subsubsection{Type I: Routing and Spectrum Assignment based on Consecutive Slots (RSACS)}

The Type I RSACS algorithm selects the path from the set of available paths. In this algorithm, whenever a new lightpath request $LR(s,d,| \Delta^{r} |)$ arrives, the algorithm traverses from the source node to its neighbouring nodes and so on, based on the available spectrum slots. If even one spectrum slot is not available through a link, the search through the link leading to this situation is terminated. The search continues through the other links to find all possible paths with at least one contiguous slot. The process continues until we reach the destination node. At the end of algorithm, we will find all possible paths from source node to the destination node along with the number of available contiguous slots. The best path among all the found paths having the required number of contiguous slots, is allocated to the lightpath request  based on the first-fit spectrum assignment as shown in Algorithm \ref{alg:type1}. The set of \textit{k} (where \textit{k} is a positive integer) possible paths are computed using \textsc{CandidatePaths()} function (Algorithm \ref{fun:cps1}). In this function, the search for the candidate paths starts from the source node \textit{`s'}. Then in each iteration, based on the number of hops, the nodes are added. In other words, with the increase in the number of iteration, the number of hops for each path in the list keeps on increasing. The search is terminated when the destination node is found for at most \textit{k} paths or no more paths are possible. The paths which have slots greater or equal to the required slots are selected. Based on the first fit, the slots are chosen from the set of slots present on the selected path. 
\begin{algorithm}
    \caption{Type I: RSACS}
    \begin{algorithmic}[1]
        \State \textbf{Input:} $G(V, E, \{ \Delta_{e} \} )$, $LR(s, d, |\Delta^{r}|, k)$
            \State $AllPath \leftarrow$ list of \textit{k} candidate paths alongwith available slots from s to d using \Call{CandidatePaths}{$G(V, E, \{ \Delta_{e}\} )$, $s$, $d$, $k$} \Comment{\textcolor{blue}{Algorithm \ref{fun:cps1}}}
            \If {$AllPath$ is not empty}
                \For {all the paths in $AllPath$ indexed by $i$}
                    \If {$ |\Call{MaxContg}{\Delta_i}| >= |\Delta^{r}|$}\Comment{\textcolor{blue}{\Call{MaxContg}{} returns the number of maximum contiguous slots available in bitmap.}}
                        \State Return $BestPath  = AllPath(i)$   \Comment{\textcolor{blue}{First one in the AllPath() which satisfy the constraint is picked. There can be other ways of choosing if more than one options are there.}}  
                    \Else 
                        \State $i++$   
                    \EndIf
                \EndFor 
                \If {$BestPath$ is empty}   
                	\State Block the request
                \EndIf            
            \Else
                \State Block the request                
            \EndIf
       
    \end{algorithmic}
    \label{alg:type1}
\end{algorithm}

Although for single-path routing, this algorithm has higher time complexity, it is still useful for multipath routing \cite{c9}, \cite{param} as the source node has the details of all the paths with spectrum slots available in them. If there are no paths with required available slots, the source node can select multiple paths whose combined available slots are greater than or equal to the required slots. Though, for the feasibility of this scenario, capability to split the transmitted signal into multiple streams should exist at the source node. We can reduce the blocking probability of the lightpath requests with this method.

\subsubsection{Type II: Routing and Spectrum Assignment based on Consecutive Slots}
Algorithm \ref{alg:type2} is second RSA algorithm. It is called Type II: Routing and Spectrum Assignment based on Consecutive Slots. It is another way of routing using consecutive spectrum slots. Each node checks the contiguous required slots with the slots present after Bit-wise ANDing of the bitmap of path from source node and the bitmap of the link from the current node to the neighbouring node. If the current node is the destination node, that path will be considered for RSA. But, if the required slots are not available from source to the current node, then no path is feasible via current node. In this method, computational time is reduced as we need to compute the candidate path only with minimum required contiguous slots. 

The role of function \textsc{CandidatePaths()} (Algorithm \ref{fun:cp}) changes slightly and it finds best candidate paths with at least $|\Delta^{r}|$ contiguous slots. In this function, the search for the candidate path originates from the source node \textit{'s'}. Then for each iteration, the number of hops for each path in the list keeps on incrementing.

\begin{algorithm}[H]
\caption{\textsc{CandidatePaths()}}
    \begin{algorithmic}[1]
        \Function{CandidatePaths}{$G(V, E, \{ \Delta_{e}\} )$, $s$, $d$, $k$}\Comment{\textcolor{blue}{First \textit{k} paths during the search returned by the function.}}
        \State $AllPath$ = [ ] \Comment{\textcolor{blue}{Contains the list of all the paths possible with slots available in them, at the end of function.}}
        \State $Path = [s]$ \Comment{\textcolor{blue}{A path is an ordered list of nodes with neighbouring nodes having a link between them. Path is initialized with one entry for source \textit{s}.}}
        \While {$Path$ is not empty}
            \State $tmp$ = [ ]
            \For {all paths in \textit{Path} indexed by $i$}
                \State $u = Path(i).end$ \Comment{\textcolor{blue}{ $u$ is the end node as of now for the $i^{th}$ path entry}}
                \State $\{ v \} = \Call{Adj}{u}) \backslash Path(i)$  \Comment{\textcolor{blue}{\Call{ADJ}{} function returns set of all neighbours of \textit{u} excluding the ones already on $Path(i)$, $\backslash$ sign indicate exclusion. All nodes in the $i^{th}$ path are excluded}}
                \If {$\{ v \}$  is not empty}
                    \For {all nodes in $\{ v \}$ indexed by $j$}
                        \State $\Delta_{i} \leftarrow \Delta_{i} \cap \Delta_{(u,v(j))}$
                        \If {$\Call{sum}{\Delta_{i}} \neq 0$} \Comment{\textcolor{blue}{\Call{sum}{} function gives the number of available slots in $Path$, $i$ with bitmap $\Delta_{i}$. ($Path(i), v(j)$) should not be considered if sum is zero.}}
                            \If {$v(j) == d $}
                                \State $AllPath = AllPath + [(Path(i), v(j));{\Delta_{i}}]$ \Comment{\textcolor{blue}{add the $[(Path(i), v(j));{\Delta_{i}}]$, as another path from $s$ to $d$ with ${\Delta_{i}}$ available slots.}}
                                \If {$size(AllPath) == k$}
                                    \State \textbf{Return:} $AllPath$
                                \EndIf
                            \Else
                                \State $tmp = tmp + [(Path(i), v(j));{\Delta_{i}}]$ \Comment{\textcolor{blue}{add the path $[(Path(i), v(j));{\Delta_{i}}]$ to $tmp$ path storage.}}
                            \EndIf
                        \EndIf
                    \EndFor
                \EndIf
            \EndFor
            \State $Path = tmp$ \Comment{\textcolor{blue}{All new paths learnt are stored in $Path$. All older values in $Path$ discarded.}}
        \EndWhile
        \State \textbf{Return:} $AllPath$
        \EndFunction
    \end{algorithmic}
    \label{fun:cps1}
\end{algorithm}

\begin{algorithm}
    \caption{Type II: RSACS}
    \begin{algorithmic}[1]
        \State \textbf{Input:} $G(V, E, \{ \Delta_{e} \} )$, $LR(s, d, |\Delta^{r}|, 1)$
        
            \State $AllPath \leftarrow$ A candidate path (k = 1) from $s$ to $d$ using \Call{CandidatePaths}{$G(V, E, \{ \Delta_{e} \} )$, $LR(s, d, |\Delta^{r}|, 1)$} \Comment{\textcolor{blue}{Algorithm \ref{fun:cp}}}
            \If {$AllPath$ is not empty}
               \State $BestPath$ = $AllPath(1)$
               \State \textbf{Return} $BestPath$
            \Else
                \State Block the request;
                
            \EndIf
        
    \end{algorithmic}
    \label{alg:type2}
\end{algorithm}

 \textsc{IsFeasible()} function (Algorithm \ref{fun:if}) checks the required number of slots $|\Delta^{r}|$ at each node, if available, then passes the information to the adjacent nodes for route discovery further. Else, the path with slots $\leq$ $|\Delta^{r}|$ is discarded from the list. The search is stopped when the destination node is found for one of the paths (\textit{k} = 1). Then, Algorithm \ref{alg:type2} checks the \textit{AllPath}. If it is not empty, then that path is used for RSA. Whereas, if the path is not found, then that lightpath request gets blocked.

This way of routing is best suited for single path routing i.e. \textit{k} = 1.

\begin{algorithm}
\caption{\textsc{CandidatePaths()} $\rightarrow$ Path with contiguous slots $\geq$ $\Delta^r$}
    \begin{algorithmic}[1]
        \Function{CandidatePaths}{$G(V, E, \{ \Delta_{e}\} )$, $LR(s, d, |\Delta^{r})|, k$}
       
        \State $AllPath$ = [ ]
        \State $Path = [s]$
        \While {$Path$ is not empty}
            \State $tmp$ = [ ]
            \For {all paths in \textit{Path} indexed by $i$}
                \State $u = Path(i).end$
                \State $\{v\} = \Call{Adj}{u}) \backslash Path(i)$ 
                \If {$\{v\} \neq \text{[]}$}
                    \For {all nodes in $\{ v \}$ indexed by $j$}
                        \State $\Delta_{i} \leftarrow \Delta_{i} \cap \Delta_{(u,v(j))}$
                        \If{\Call{IsFeasible}{$(\Delta_i, |\Delta^{r}|$} $==\textsc{True}$}
                                \If {$v(j) == d $}
                                    \State $AllPath = AllPath + [Path(i), v(j)]$
                                    \If {$size(AllPath) == k$}
                                        \State \textbf{Return:} $AllPath$
                                    \EndIf
                                    
                                \Else
                                    \State $tmp = tmp + [Path(i), v(j)]$
                                \EndIf
                            
                            \EndIf
                    \EndFor
                \EndIf
            \EndFor
            \State $Path = tmp$
        \EndWhile
        \State \textbf{Return:} $AllPath$ 
        \EndFunction
    \end{algorithmic}
    \label{fun:cp}
\end{algorithm}

\begin{algorithm}
\caption{\textsc{IsFeasible()}}
    \begin{algorithmic}[1]
    \Function {IsFeasible}{$(\Delta_a, |\Delta_{b}|)$}
        \Comment{\textcolor{blue}{\Call{IsFeasible}{} function checks the $|\Delta_{b}|$ number of contiguous slots are available in the bitmap $\Delta_a$}}
        \If {$|\Delta_{b}|$ can be accommodated in $\Delta_a$}
            \State \textbf{Return:} \textsc{True}
        \Else
        	\State \textbf{Return:} \textsc{False}
        \EndIf
    \EndFunction
    \end{algorithmic}
    \label{fun:if}
\end{algorithm}

\subsubsection{Type III: Routing and Spectrum Assignment based on Consecutive Slots and Shortest Path}
The Algorithm \ref{alg:type3} is Type III  Routing and Spectrum Assignment based on Consecutive Slots with additional Shortest Path Constraint. The method for computing path is same as used for Type II Algorithm except now instead of single path, multiple candidate paths are maintained from the source node \textit{s} to destination \textit{d} using \textsc{CandidatePaths} function (Algorithm \ref{fun:cp}). In this function, the search for the candidate path originates from the source node \textit{'s'}. Then for each iteration, the number of hops for each path in the list keeps on incrementing. Also, the paths with maximum contiguous slots $\leq$ $|\Delta^{r}|$  are discarded. The search is stopped when the destination node is found for at most \textit{k} paths. Based on the distance, the shortest path from \textit{AllPath} is chosen for Routing and Spectrum Assignment. If the path is not found, then that lightpath request gets blocked.

This method is the reverse process of finding k-shortest path first and then finding one with the required contiguous slots. Here we find path with required slots, and then choose the shortest one among them. This is also used for single path routing if the value of k is 1. 

\begin{algorithm}
    \caption{Type III: RSACSSP}
    \begin{algorithmic}[1]
        \State \textbf{Input:} $G(V, E, \{ \Delta_{e} \} )$, $LR(s, d, |\Delta^{r}|, k)$
                 
        \State $AllPath \leftarrow$ list of candidate paths from s to d using \Call{CandidatePaths}{$G(V, E, \{ \Delta_{e} \} )$, $LR(s, d, | \Delta^{r} |, k)$} \Comment{\textcolor{blue}{Algorithm \ref{fun:cp}}}
            \If {$AllPath$ is not empty}
                \For {all the candidate paths in $AllPath$ indexed by $i$}
                	\State Select the shortest path (distance in km) from all paths listed in $AllPath$ and store in \textit{BestPath}
                	\State \textbf{Return:} \textit{BestPath}
                \EndFor
        \Else
            \State The request is marked blocked.
            
        \EndIf
        
    \end{algorithmic}
    \label{alg:type3}
\end{algorithm}

\section{Numerical Results}

The \textit{k}-shortest path (\textit{k}-sp) algorithm finds the route on the basis of path containing least number of hops or shortest distance (km or miles). After that it checks whether the required slots are available or not on the found paths. This method might require extra computation. Therefore, we proposed algorithms in which the routes are found on the basis of contiguous and continuous spectrum  slots availability. The performance of the proposed algorithms is compared with the \textit{k}-shortest path algorithm and shortest path algorithm. 

\begin{figure}
\begin{subfigure}{0.4\textwidth}
    \centering
    \includegraphics[width=\linewidth]{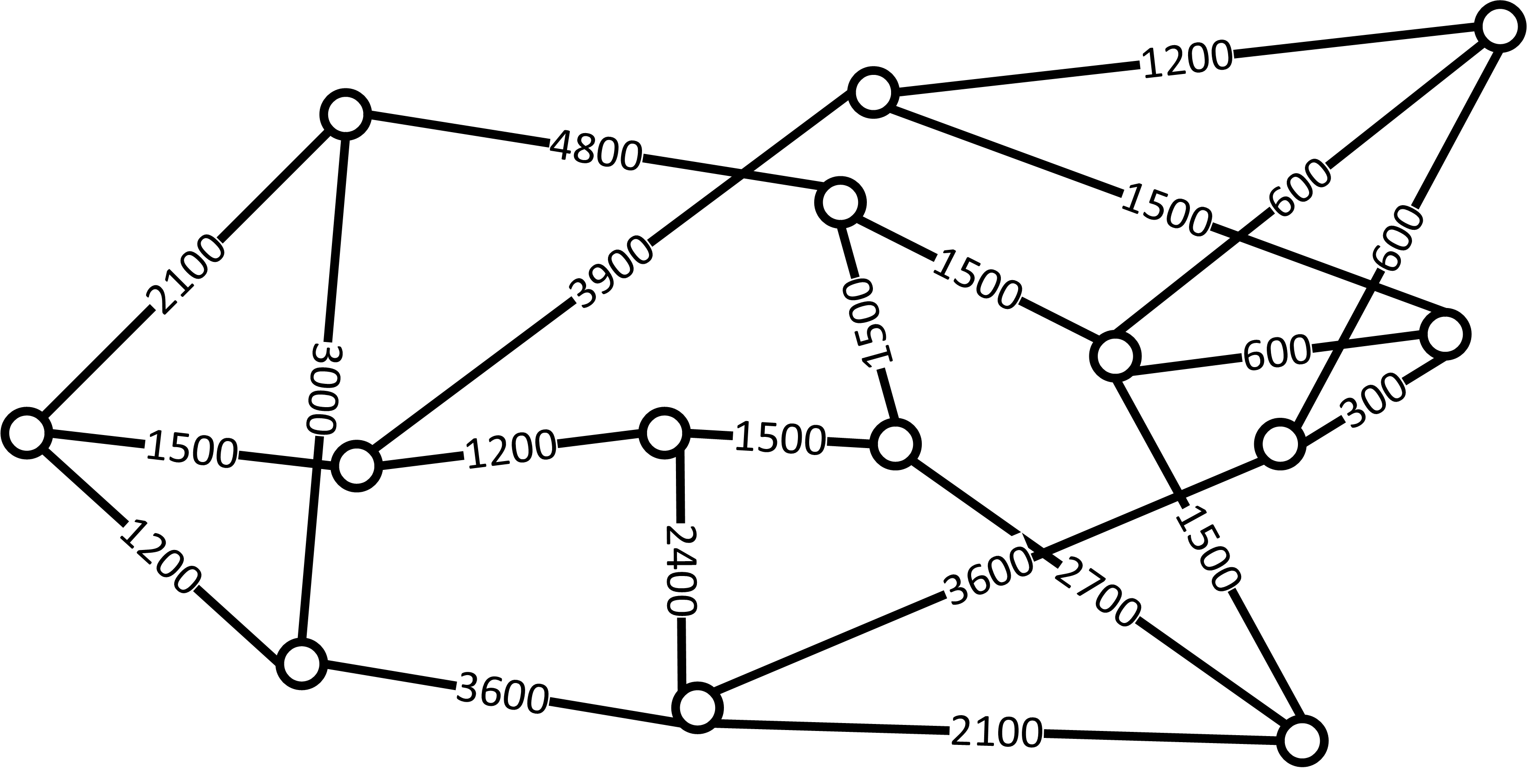}
    \caption{14 nodes, 22 links NSFNET.}
    \label{fig:nsfnet}
\end{subfigure}
\begin{subfigure}{0.4\textwidth}
    \centering
    \includegraphics[width=\linewidth]{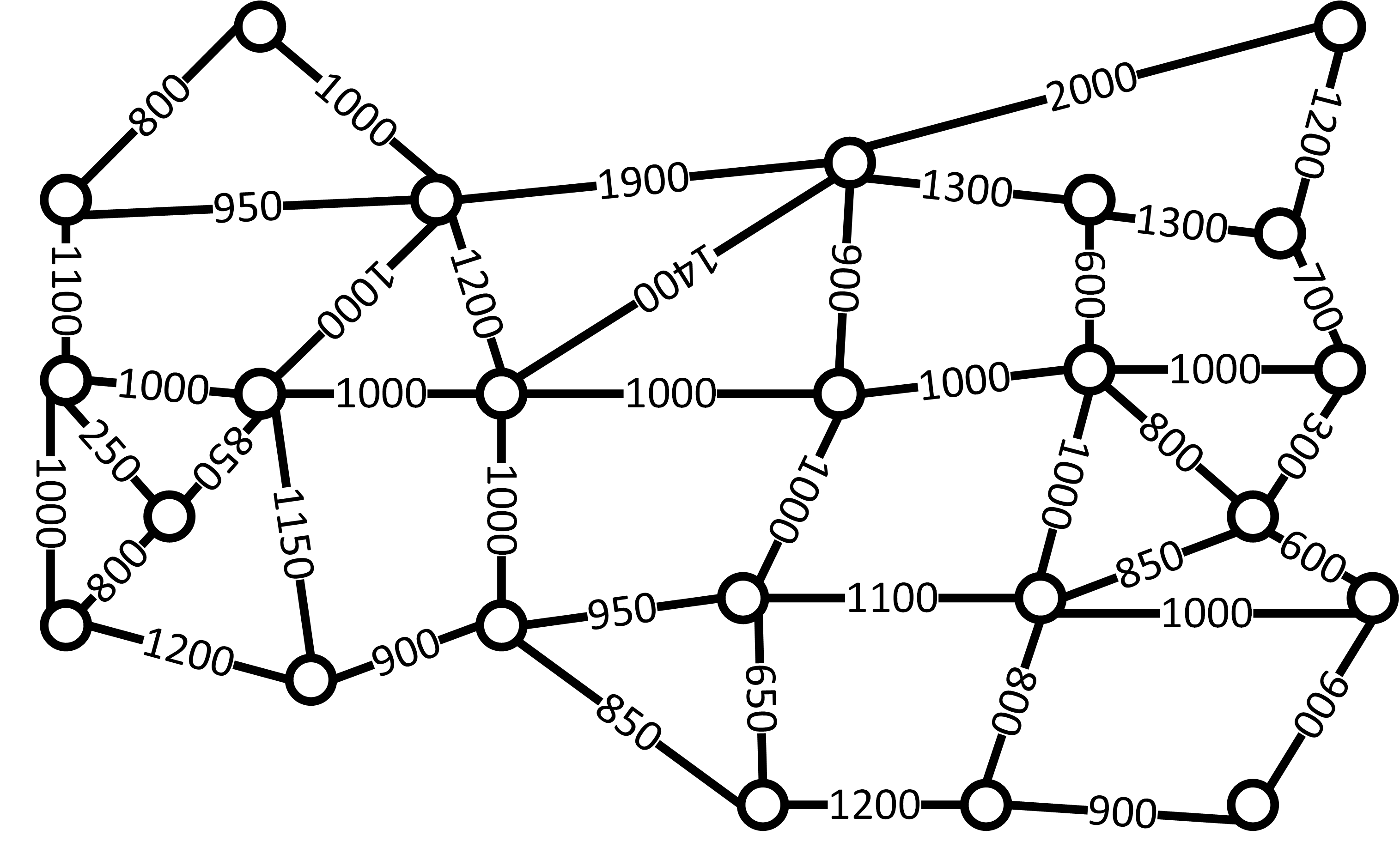}
    \caption{24 nodes, 43 links USNET.}
    \label{fig:usnet}
\end{subfigure}
\caption{Networks Topologies used.}
\label{fig:e}
\end{figure}

\subsection{Network Settings}
To evaluate the efficacy of our proposed algorithms, we operated a set of simulation experiments using MATLAB R2019b. The simulations are done for 2,00,000 requests for multiple iterations. The performance of proposed algorithms - Type I: RSACS, Type II: RSACS and Type III: RSACSSP in Flexigrid Optical Networks, are evaluated on the 14-nodes 22-links NSFNET with an average nodal degree\footnote{nodal degree, $n_d = \dfrac{2E}{V}$} 3.0, and 24 nodes 43 links USNET with an average nodal degree 3.5 as shown in Figure \ref{fig:e}. We assume the fiber bandwidth in the frequency range of C-band to be 4 THz on each link of the network. Using O-OFDM technology, the whole bandwidth is divided into 12.5 GHz parallel channels. Therefore, there are 320 spectrum slots on each link of the network, calculated using eq. \ref{eqn0}. The traffic demands for all the lightpath requests on each node pair are uniformly distributed. The bandwidth required for each lightpath is chosen randomly between 1 and B, where different 'B' values are used as a parameter in simulations. In this paper, the values of B are 100 Gbps\footnote{8 spectrum slots, if the grid size is 12.5 GHz} and 200 Gbps\footnote{16 spectrum slots, if the grid size is 12.5 GHz}. For spectrum allocation, we also considered an additional guard band (GB). The size of GB is considered to be 10 GHz. 

The use of static traffic for simulation does not show the effectiveness of the proposed algorithms. Hence, lightpath requests were  generated dynamically, i.e., we considered dynamic traffic scenario. The incoming lightpaths can be set up and released upon request. These are equivalent to setting up and releasing circuits in circuit-switched networks. The incoming lightpath requests arrive with an exponentially distributed inter-arrival time with the average of $\dfrac{1}{\lambda}$ seconds. Each connection is maintained for exponentially distributed holding time with average of $\dfrac{1}{\mu}$ seconds before being released. The offered load ($\rho$) in Erlang (E) is given by

\begin{equation}
\rho = \dfrac{\dfrac{1}{\mu}}{\dfrac{1}{\lambda}} = \dfrac{\lambda}{\mu}
\end{equation} 
The values of offered load per node ($\rho$) is considered to be 2 to 30 Erlangs.

The performance of the \textit{k}-shortest path (\textit{k}-sp) algorithm depends on the number of shortest paths \textit{k} used for pathfinding. On the other hand, the pathfinding in our proposed algorithms additionally depends on the availability of the spectrum slots. The value of \textit{k} chosen for pathfinding is 10 for \textit{k}-sp, Type I and Type III algorithms. At the same time, the shortest path and  Type II algorithms are independent of the value \textit{k}. The performance parameters are estimated based on the observations made during the steady-state condition (which is observed to happen after approximately three times the average holding time, i.e., $3*\dfrac{1}{\mu}$).

\subsection{Performance Metrics} 
The performance of the proposed algorithms is evaluated on the basis of blocking probability of the incoming requests, blocking probability of the incoming required slots, and the spectrum utilization with the gradual increments in demand rate (in Gbps): 
    \begin{itemize}
        \item \textbf{Blocking Probability}: It is defined as the ratio of total number of blocked connections to the total number of arrived connections.
        \item \textbf{Bandwidth Blocking Probability}: It is defined as the ratio of the total amount of incoming bandwidth or slots which are blocked, to the total amount of bandwidth or slots required by all the incoming connections.
        \item \textbf{Spectrum Utilization}: The ratio of total bandwidth (or slots) used to the total bandwidth (or slots) in the spectrum.
    \end{itemize}

\subsection{Simulation Results}
\begin{figure}
    \centering
    \includegraphics[width=0.8\linewidth]{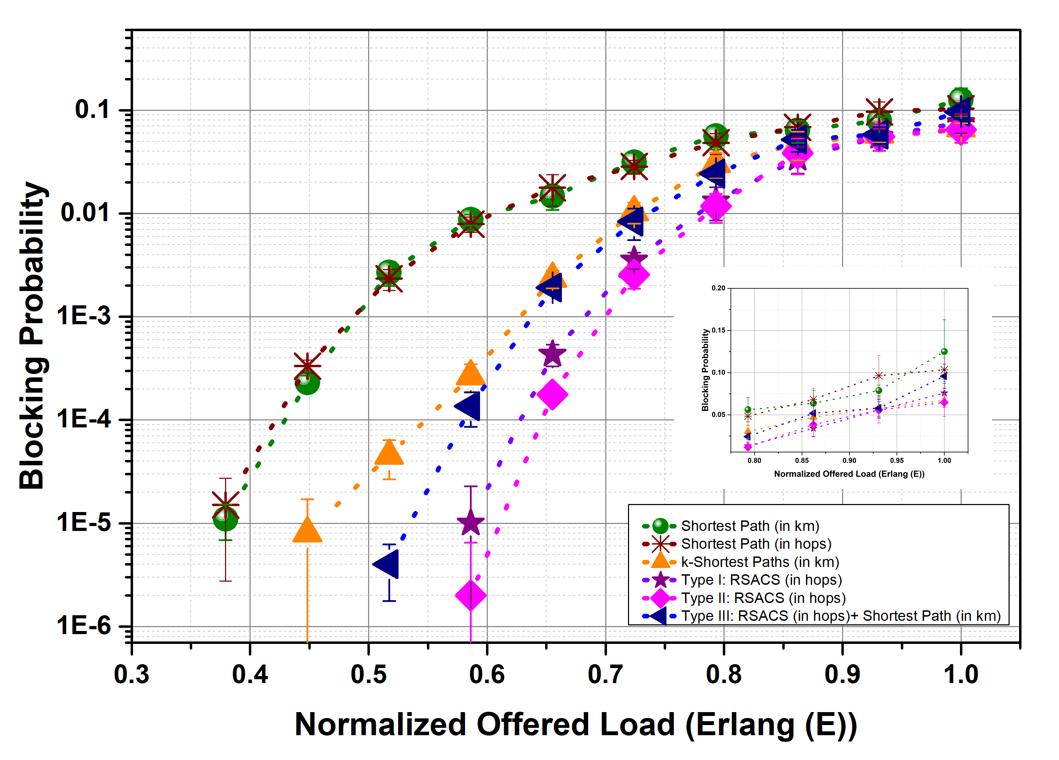}
    \caption{Blocking Probability Vs Offered Loads for NSFNET with a demand rate of 100 Gbps.}
    \label{fig:bpk10d100}
\end{figure}

\begin{figure}
    \centering
    \includegraphics[width=0.8\linewidth]{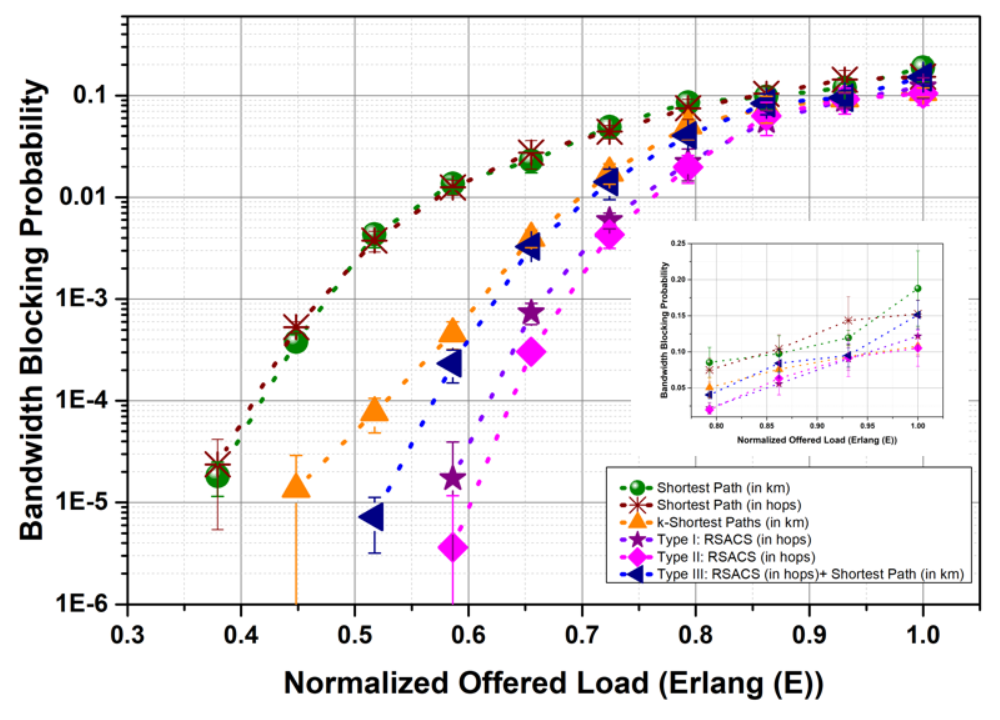}
    \caption{Bandwidth Blocking Probability Vs Offered Loads for NSFNET with a demand rate of 100 Gbps.}
    \label{fig:bbpk10d100}
\end{figure}

\begin{figure}
    \centering
    \includegraphics[width=0.8\linewidth]{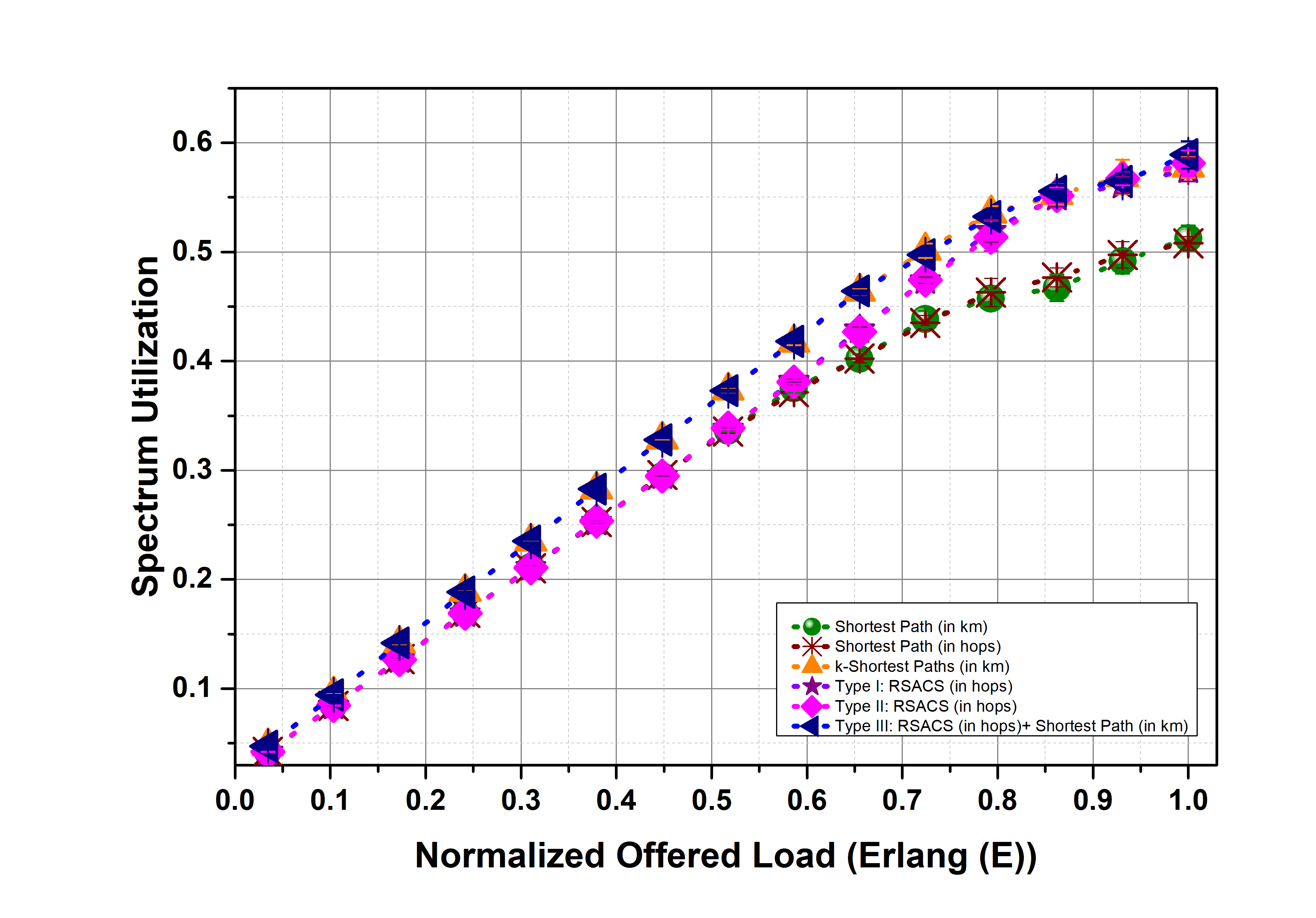}
    \caption{Spectrum Utilization Vs Offered Loads for NSFNET with a demand rate of 100 Gbps.}
    \label{fig:suk10d100}
\end{figure}

\begin{figure}
    \centering
    \includegraphics[width=0.8\linewidth]{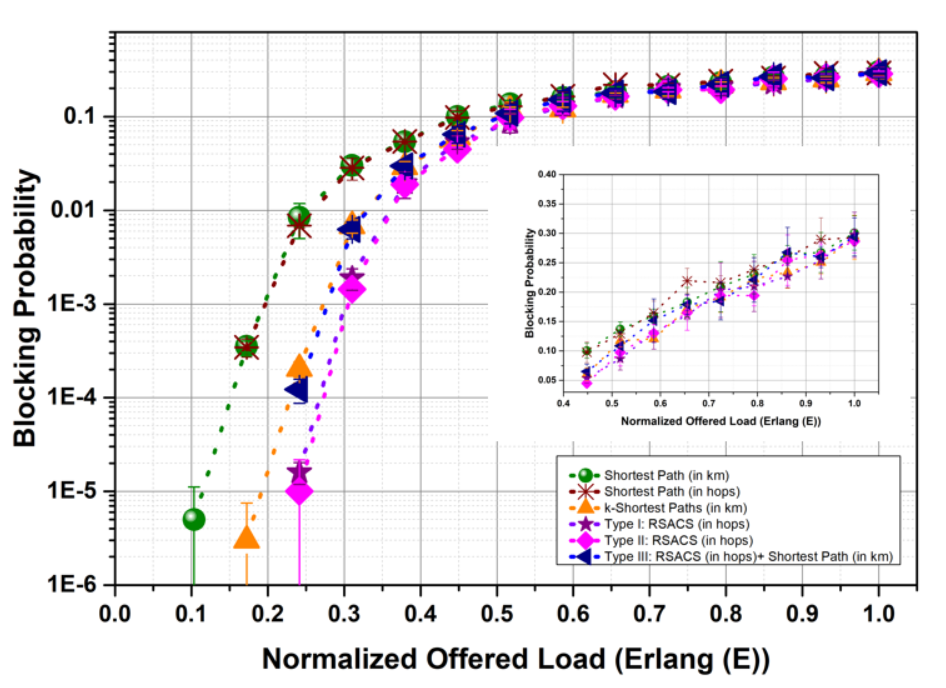}
    \caption{Blocking Probability Vs Offered Loads for NSFNET with a demand rate of 200 Gbps.}
    \label{fig:bpk10d200}
\end{figure}

\begin{figure}
    \centering
    \includegraphics[width=0.8\linewidth]{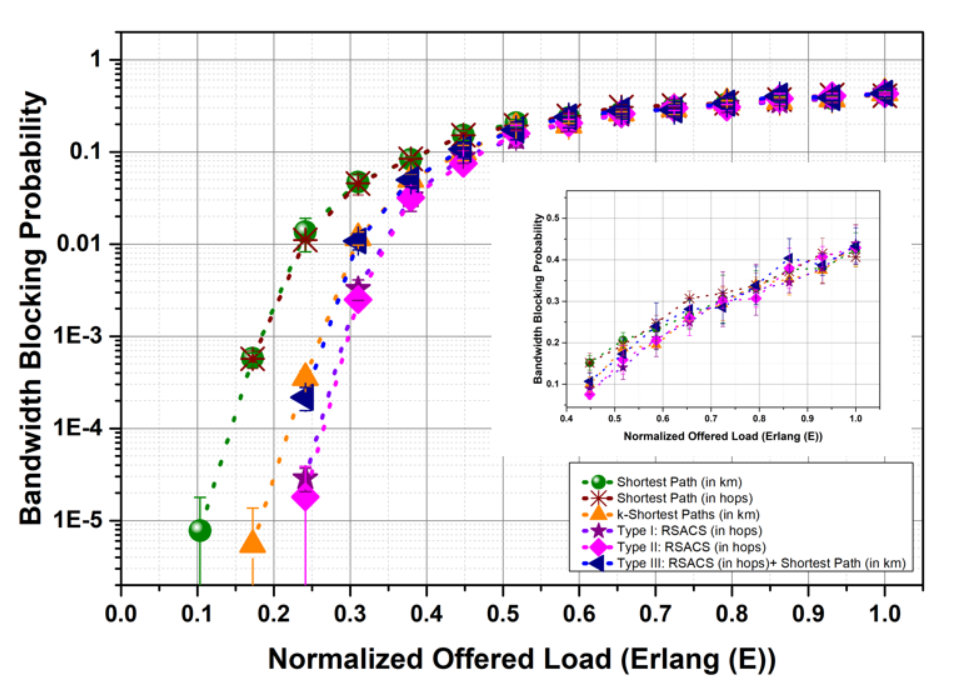}
    \caption{Bandwidth Blocking Probability Vs Offered Loads for NSFNET with a demand rate of 200 Gbps.}
    \label{fig:bbpk10d200}
\end{figure}

\begin{figure}
    \centering
    \includegraphics[width=0.8\linewidth]{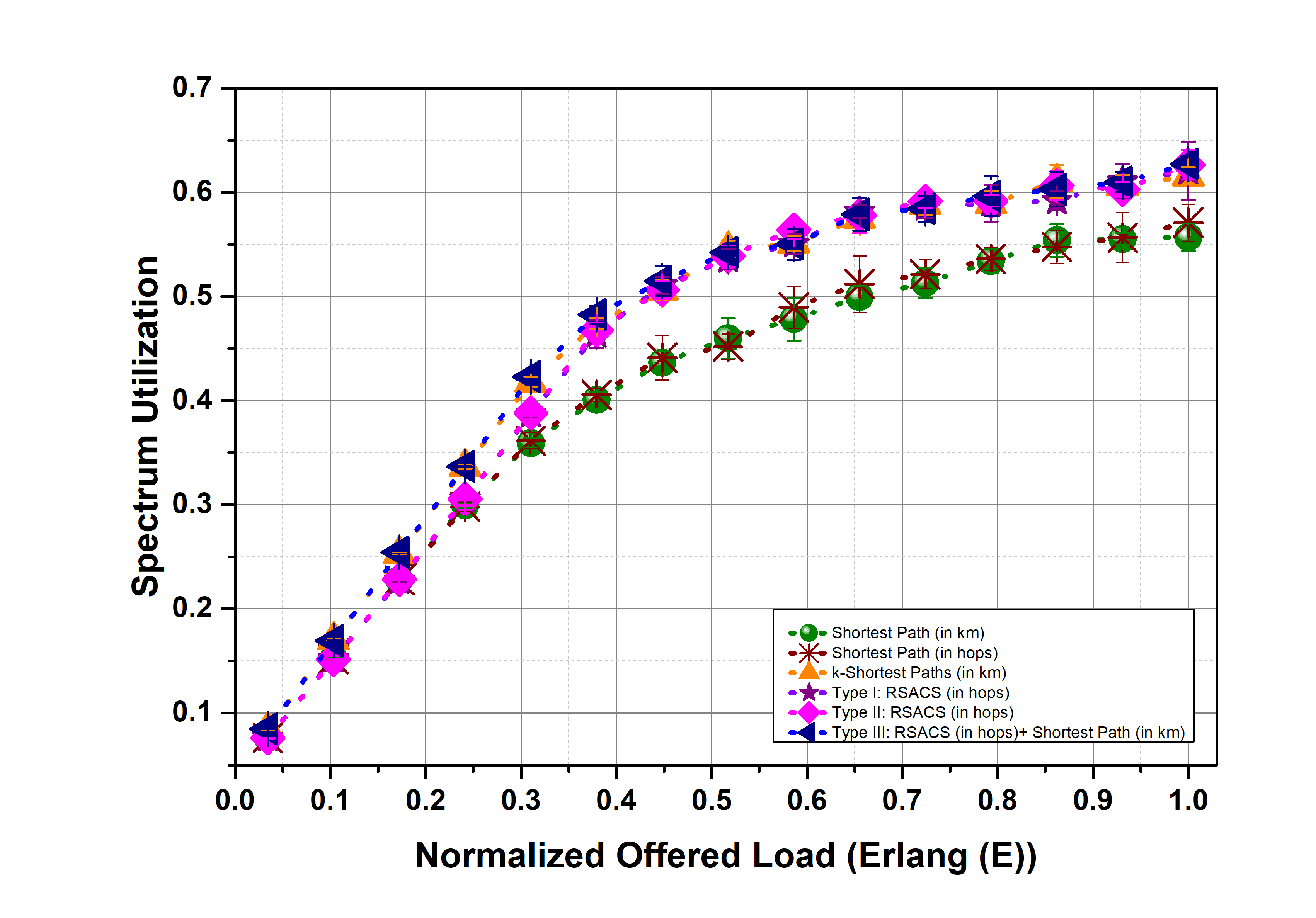}
    \caption{Spectrum Utilization Vs Offered Loads for NSFNET with a demand rate of 200 Gbps.}
    \label{fig:suk10d200}
\end{figure}

\begin{figure}
    \centering
    \includegraphics[width=0.8\linewidth]{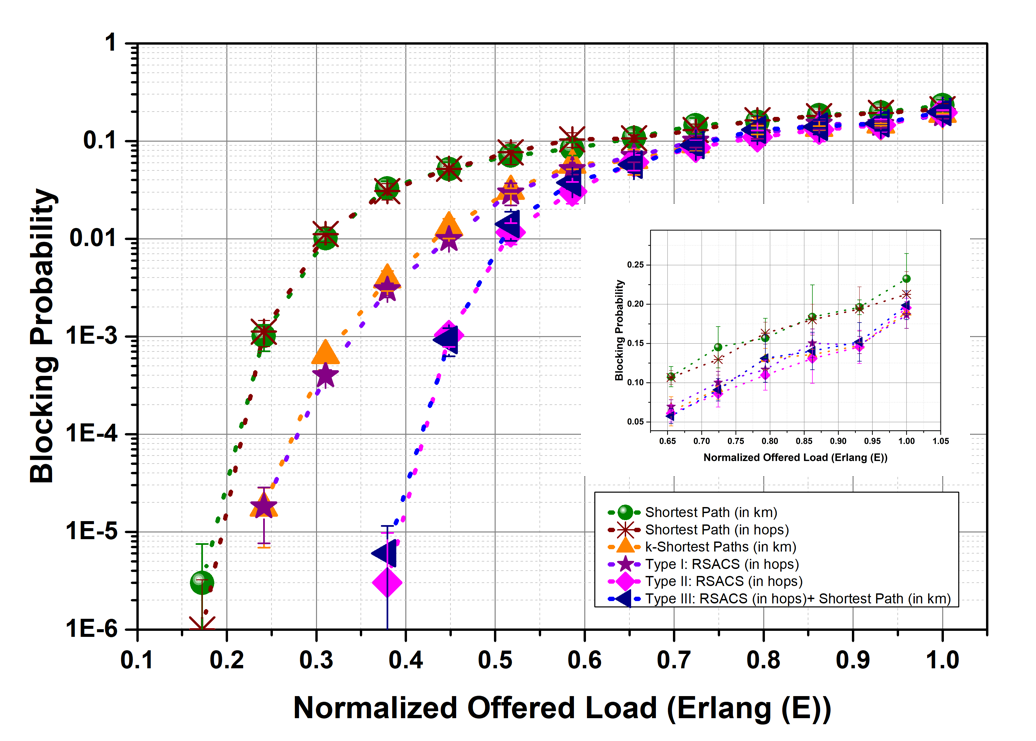}
    \caption{Blocking Probability Vs Offered Loads for USNET with a demand rate of 100 Gbps.}
    \label{fig:bpuk10d100}
\end{figure}

\begin{figure}
    \centering
    \includegraphics[width=0.8\linewidth]{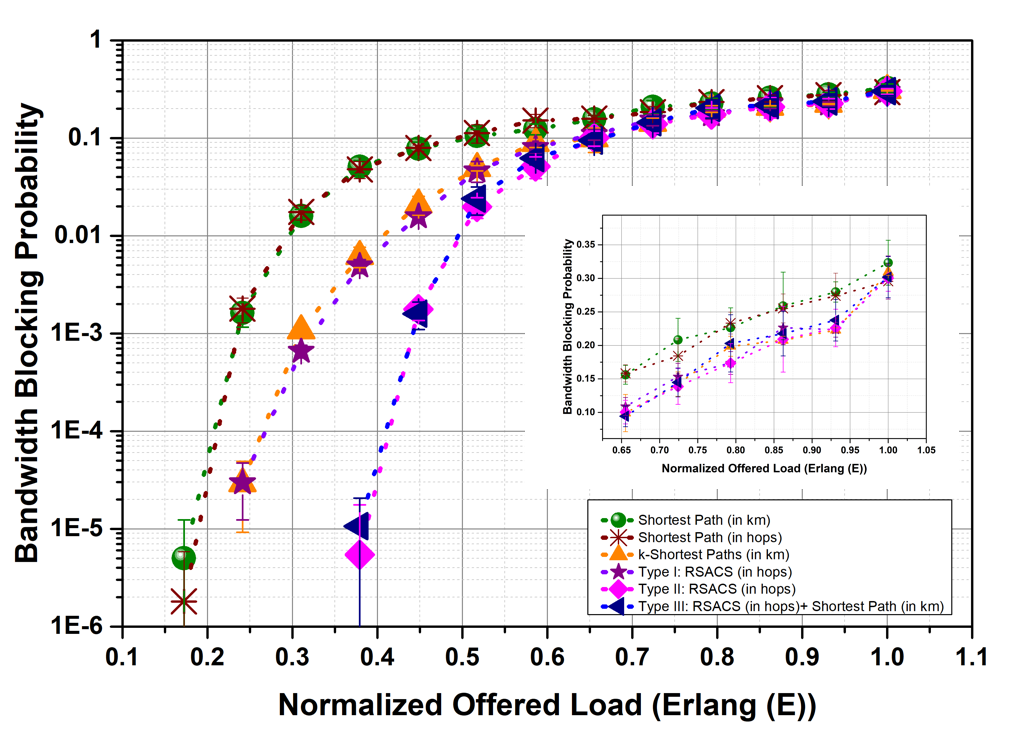}
    \caption{Bandwidth Blocking Probability Vs Offered Loads for USNET with a demand rate of 100 Gbps.}
    \label{fig:bbpuk10d100}
\end{figure}

\begin{figure}
    \centering
    \includegraphics[width=0.8\linewidth]{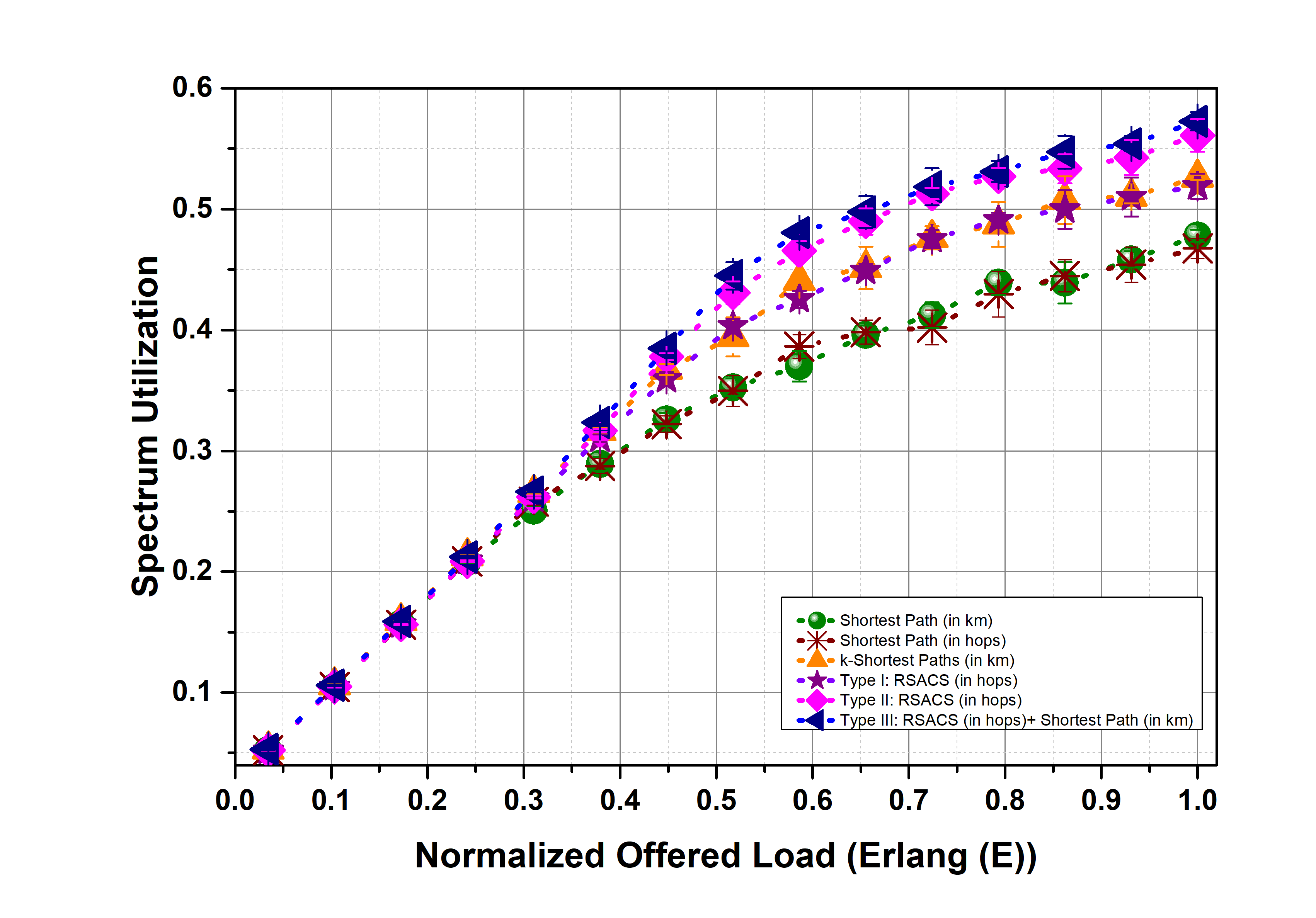}
    \caption{Spectrum Utilization Vs Offered Loads for USNET with a demand rate of 100 Gbps.}
    \label{fig:suuk10d100}
\end{figure}

\begin{figure}[h]
    \centering
    \includegraphics[width=0.9\linewidth]{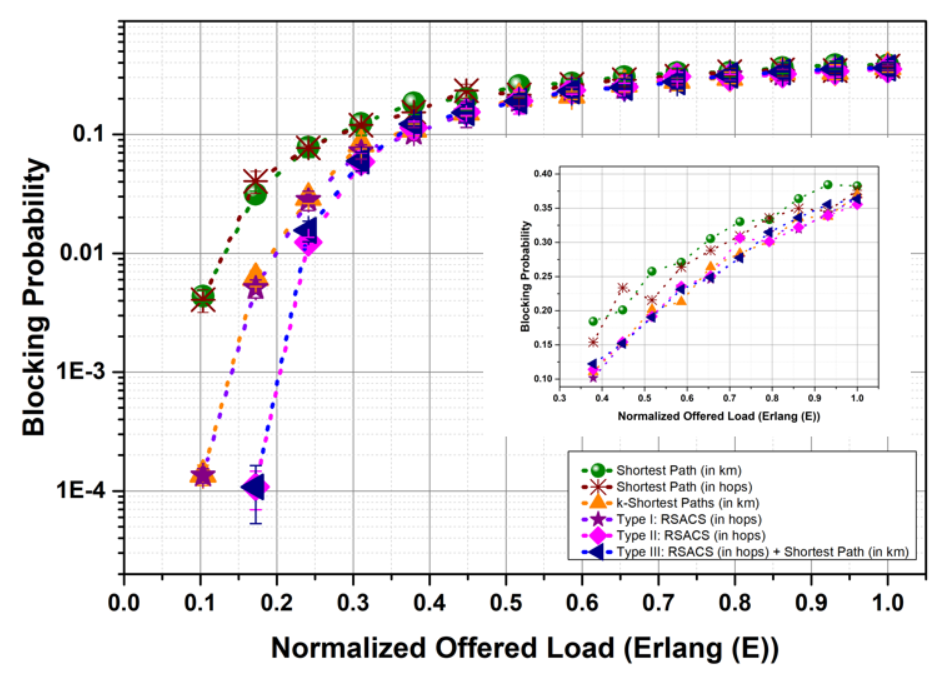}
    \caption{Blocking Probability Vs Offered Loads for USNET with a demand rate of 200 Gbps.}
    \label{fig:bpuk10d200}
\end{figure}

\begin{figure}
    \centering
    \includegraphics[width=0.9\linewidth]{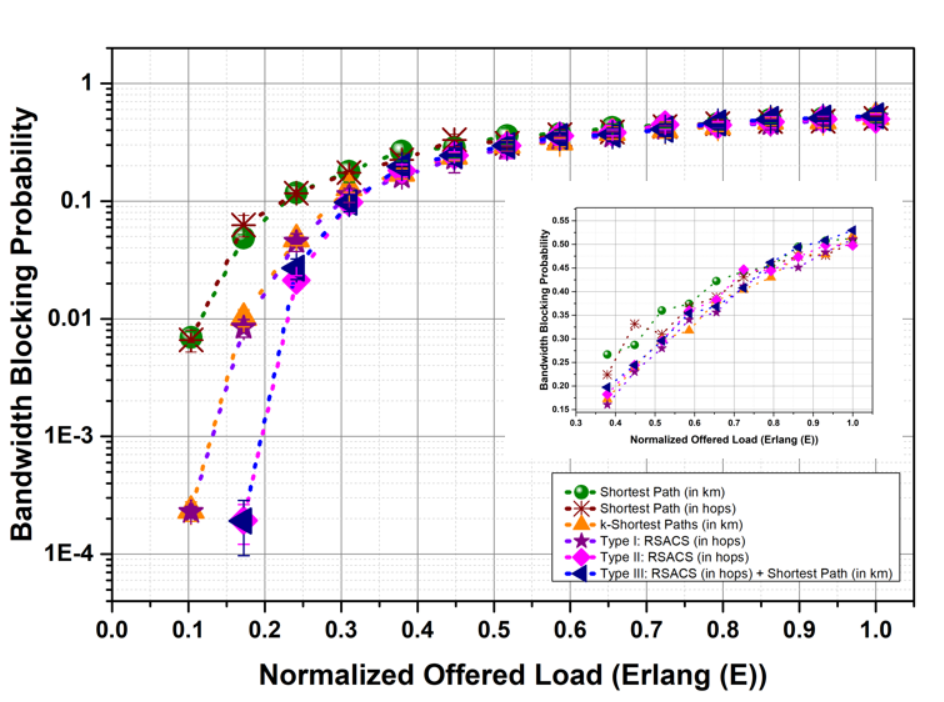}
    \caption{Bandwidth Blocking Probability Vs Offered Loads for USNET with a demand rate of 200 Gbps.}
    \label{fig:bbpuk10d200}
\end{figure}

\begin{figure}
    \centering
    \includegraphics[width=0.9\linewidth]{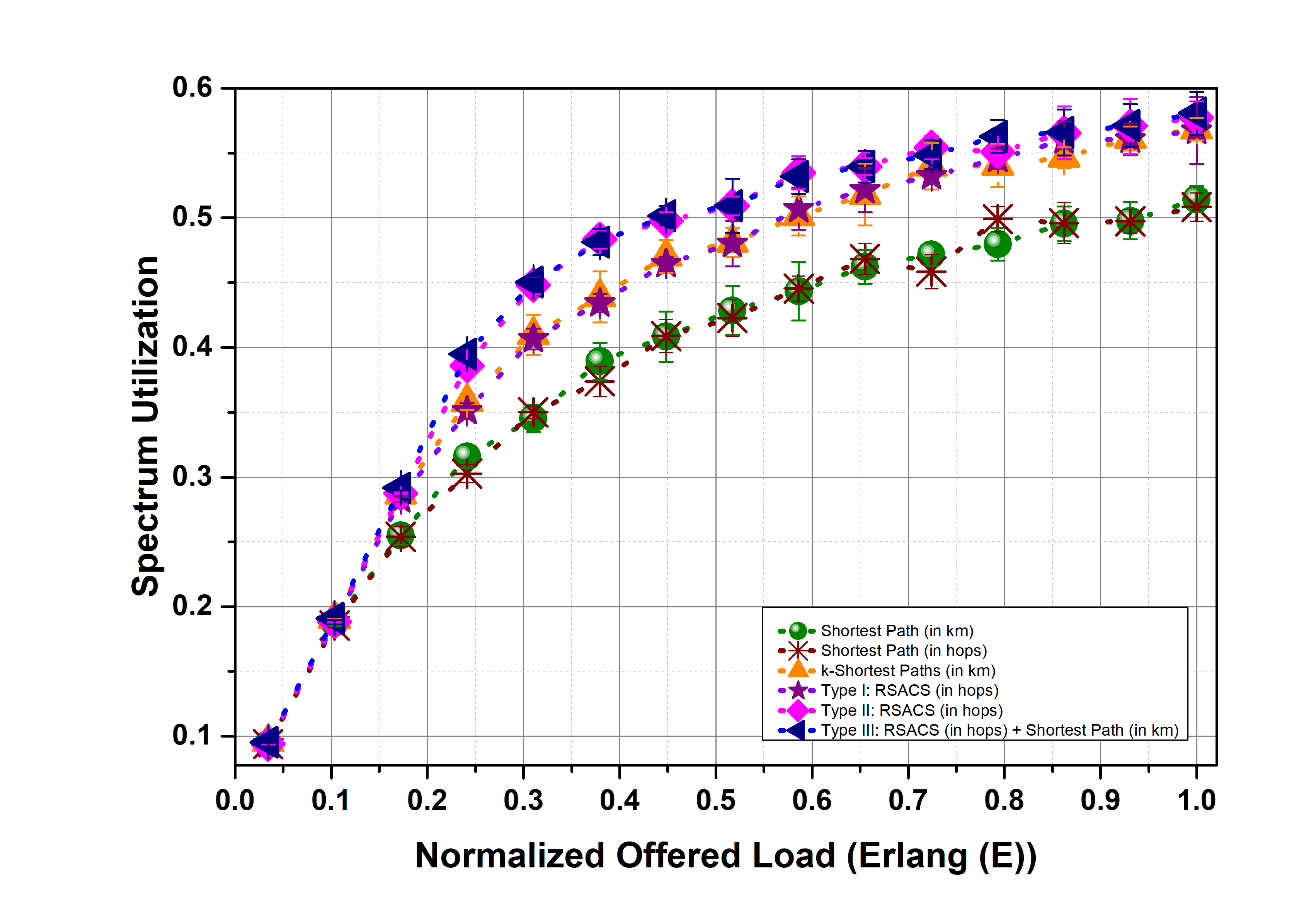}
    \caption{Spectrum Utilization Vs Offered Loads for USNET with a demand rate of 200 Gbps.}
    \label{fig:suuk10d200}
\end{figure}

The path finding algorithms used for comparison with the proposed algorithms are shortest path (in km), shortest path (in hops),  and \textit{k}-shortest path (in km) algorithms. Figures \ref{fig:bpk10d100}, \ref{fig:bpk10d200}, \ref{fig:bpuk10d100}, and \ref{fig:bpuk10d200} show the performance of algorithms in terms of blocking probability of the incoming lightpath requests; whereas figures \ref{fig:bbpk10d100}, \ref{fig:bbpk10d200}, \ref{fig:bbpuk10d100}, and \ref{fig:bbpuk10d200} are for the blocking probability of incoming spectrum slots, and figures \ref{fig:suk10d100}, \ref{fig:suk10d200}, \ref{fig:suuk10d100}, and \ref{fig:suuk10d200} are for the utilization of the spectrum slots. In all the cases, the spectrum utilization performance can be observed after the blocking probability of 0.01 as before 0.01 blocking probability, spectrum utilization is same due under-utilization of the spectrum slots. After 0.01 blocking probability, one can see the change in the performance due to fragmentation. 

Figures \ref{fig:bpk10d100}, \ref{fig:bbpk10d100} and \ref{fig:suk10d100} are for NSFNET network where the value of \textit{k} is 10 and the incoming demand rate is 100 Gbps. 

Figures \ref{fig:bpk10d100} and \ref{fig:bbpk10d100} plot the blocking probability and bandwidth blocking probability of six pathfinding techniques with first-fit spectrum allocation policy. Intuitively, as the value of the offered load increases, the blocking probability increases, as shown in figures \ref{fig:bpk10d100}  and \ref{fig:bbpk10d100}. The performance of consecutive slots-based pathfinding algorithms is better than the shortest path algorithms. Additionally, for low loads, Type II algorithms outperform other algorithms. Also, the Type III and \textit{k}-shortest path algorithms are almost the same after $60\%$ of the offered load, except in Type III, the adaptive parameter is used first and then static parameter for routing. The performance of the Type III algorithm is better than the \textit{k}-shortest path algorithm for lower loads. Intuitively, the shortest path algorithm finds the route from source to destination based on the distance, not the spectrum slots available. Therefore, if the spectrum slots are not available at the time of spectrum assignment, then the path gets blocked. In this case, other paths might have required spectrum slots; therefore Type-III algorithm is expected to perform better.

The Type II algorithm finds the path based on required spectrum slots. The path gets blocked at the routing time if no path is available with the required spectrum slots. In this, all the possible paths with required spectrum slots are checked until the destination node is found. Therefore, the Type-II algorithm's time complexity for the worst case scenario is higher than the shortest path algorithm, as given in Table \ref{table:II}. Type II and shortest path algorithms are independent of the value of \textit{k}. However, the performance of Type II is far better than the shortest path for all load conditions. Shortest path algorithm has the worst performance among all the algorithms.%Nevertheless, after the $0.45$ offered load, the blocking performance is almost the same.
\begin{table*}[h]
	
    \begin{center}
        \centering
        \resizebox{\textwidth}{!}{%
        \begin{tabular}{|p{4.5cm}|p{4cm}|p{4cm}|p{4cm}|p{4cm}|p{4cm}|}
            \hline
            %\centering
            \textbf{\large Parameters} & \textbf{\large Shortest Path} & \textbf{\large \textit{k}-Shortest Path}& \textbf{\large Type I}& \textbf{\large Type II}& \textbf{\large Type III} \\
            \hline
            \large \textbf{Path finding} &  \large On the basis of distance or hops & \large On the basis of distance or hops & \large On the basis of available slots& \large On the basis of available slots & \large On the basis of available slots and distance\\
            \hline
            \large \textbf{Routing and Spectrum Assignment Problem} &  \large Separate problem & \large Separate problem & \large Joint problem & \large Joint problem & \large Joint problem\\
            \hline
            \large \textbf{Time Complexity} &\large \textit{O(V.E)+O(V.E.$\Delta$)} &  \large \textit{O(k.V.E)+O(V.E.$\Delta$)}& \large \textit{O(k.V.E.$\Delta$)}& \large \textit{O(V.E.$\Delta$)}& \large \textit{O(k.V.E.$\Delta$)}\\
            \hline
            \large \textbf{Blocking of the request after path finding} & \large If the required spectrum is not available, request can be blocked  & \large If the required spectrum is not available, request can be blocked  & \large If the required spectrum is not available, request can be blocked   & \large No blocking & \large No blocking\\
            \hline
             
        \end{tabular}}
        \caption{Comparison of Algorithms used for Routing and Spectrum Assignment. Here, V is the set of vertices, E is the set of edges, $\Delta$ is the set of spectrum slots, and \textit{k} is the maximum number of paths to be computed by RSA.}
        \label{table:II}
    \end{center}
    
\end{table*}

Figure \ref{fig:suk10d100} compares the performance in terms of spectrum utilization. The blocking probability of the connections for lower loads is almost negligible; therefore, the spectrum remains underutilized. But as the offered load values increase, there is a higher spectrum utilization for all the algorithms except shortest path algorithms. Based on the observations, higher blocking of connections and lower spectrum utilization for the shortest path is due to the unavailability of the contiguous spectrum slots in a fragmented state. In contrast, for other cases, the fragmentation is lower. 

Figures \ref{fig:bpk10d200}, \ref{fig:bbpk10d200} and \ref{fig:suk10d200} plots are for NSFNET network where the value of \textit{k} is 10 and the maximum incoming demand rate is 200 Gbps. For higher demand rates, the blocking probability keeps increasing as the availability of required spectrum slots becomes less than that for the case of 100 Gbps. The comparison is similar to that for the demand rate of 100 Gbps. The spectrum performance is also similar, i.e. the spectrum utilization of the shortest path algorithms is worst as the offered load per node increases. Whereas, other four algorithms' spectrum utilization performance is almost the same. We can notice the spectrum utilization performance due to fragmentation for lower loads, i.e., around $30\%$.

Alongside NSFNET, we used USNET network to determine the performance trends with topology change. Figures \ref{fig:bpuk10d100}, \ref{fig:bbpuk10d100} and \ref{fig:suuk10d100} are for USNET network where the value of \textit{k} is 10 and the incoming demand rate is 100 Gbps. This observation is for large networks with 24 nodes and 43 links. The blocking performance of the lightpath requests and spectrum slots changes as compared to NSFNET. In USNET also the Type II and Type III algorithm outperforms the other strategies. In USNET again the shortest path algorithms perform worst in terms of blocking probability and utilization of the spectrum slots. 

Now, as the network diameter changes, we can also observe the performance in the utilization of the spectrum slots for the other four algorithms. Type II and Type III have the higher spectrum utilization for the higher offered loads than Type I and $k$-shortest path. 

Figures \ref{fig:bpuk10d200}, \ref{fig:bbpuk10d200} and \ref{fig:suuk10d200} are for USNET network where the value of \textit{k} is 10 and the size of incoming demand rate is 200 Gbps. The blocking probability and spectrum utilization performance are almost the same. Except in such configurations, the various performance parameters keep deteriorating for lower loads due to an increase in the size of the incoming demand rate. Also, the blocking probability performance converges for higher loads for all the algorithms.

Another set of observations are presented in Table \ref{table:II}. The time complexity for the worst-case scenario is lowest for Shortest Path. Whereas time complexity for $k$-Shortest Path and Type II are comparable depending upon the value of k. Because if $k$ = 1, then the $k$-shortest path converges to the shortest path. However, if the value of $k$ is too high, then time complexity keeps worsening. Even higher than Type II. Therefore, the time complexity for the worst-case scenario is in the following order Shortest Path $<$ $k$-Shortest Path, Type II $<$ Type I, Type III.
%In all the above cases, the spectrum utilization performance can be observed after blocking probability of 0.01. Therefore, the spectrum utilization before 0.01 blocking probability is same due under-utilization of the spectrum slots. Whereas the the spectrum utilization after 0.01 blocking probability one can see the change in the performance due to fragmentation.
The Type II algorithm has better performance in blocking probability, spectrum utilization and time complexity for the worst-case scenario. Also, we know a priori blocking, i.e., at the routing time, so there is no need for spectrum assignment in this algorithm. Therefore, it can used to lower the blocking of the lightpath requests. However, eliminating fragmentation is impossible; therefore, one can use the de-fragmentation strategy in addition to Type II algorithm, instead with the $k$-shortest path such that the traffic is disrupted fewer times. 

%One can check the performance of the Type-II algorithm for multi-band optical transmissions. As the combination of C, L and S-band can provide high capacity as compared to standalone C band \cite{MB}. Hence, the convergence of the algorithms can be obtained at a higher load per node.      

\section{Conclusion}

The Routing and Spectrum Assignment is a tedious process in Flexible grid Optical Networks. The main reason is dynamic arrivals and departures of the lightpath requests and contiguity constraint, resulting in fragmentation within the spectrum of the network. In this paper, instead of using static parameters for RSA, we used adaptive parameters, i.e., consecutive spectrum slots for routing. The consecutive spectrum slots keep changing with the network conditions. We performed the detailed analysis under different demand rates for NSFNET and USNET. The performance of the proposed algorithms is better than the existing strategies in terms of lower blocking of the lightpath requests. One of the reasons is the lower fragmentation of the spectrum slots. We performed a detailed analysis of the proposed algorithms. The Type II algorithm has lower blocking probability, higher spectrum utilization and lower time complexity.

\end{document}